\documentclass{cambridge6A}
  \usepackage{graphicx}

\newcommand{\be}{\begin{equation}}
\newcommand{\ee}{\end{equation}}

\begin{document}

\chapterauthor{Harvey S. Reall}

\chapter{Algebraically special solutions in higher dimensions}

\contributor{Harvey S. Reall \affiliation{Department of Applied Mathematics and Theoretical Physics, Cambridge University}} 

\section{Introduction}

At any point of a 4d spacetime, the Weyl tensor (if non-zero) defines four Principal Null Directions (PNDs) satisfying
\be
\label{pnd}
\ell_{[\mu} C_{\nu]\rho\sigma[\tau} \ell_{\lambda]} \ell^\rho \ell^\sigma =0.
\ee
The Weyl tensor can be classified according to whether any of these null directions coincide. In general, the null directions are distinct and the Weyl tensor is said to be algebraically general. However, if two or more PNDs coincide then the Weyl tensor is algebraically special. An algebraically special {\it spacetime} is one whose Weyl tensor is everywhere algebraically special. 

The algebraically special property played an important role at several points in the history of black hole physics. The most important non-trivial solution of General Relativity is the Kerr solution. The title of Kerr's paper "Gravitational field of a spinning mass as an example of algebraically special metrics" \cite{Kerr:1963ud} reveals the motivation for Kerr's work: he was interested in finding new solutions with algebraically special Weyl tensor. Assuming that the metric is algebraically special leads to a simplification of the Einstein equation, which helped Kerr find his solution.

Once the special importance of the Kerr solution came to be appreciated, the question of its classical stability was raised. Answering this question became possible only after Teukolsky exploited the algebraically special property to decouple the equations governing linearized metric perturbations of the Kerr solution and reduce them to a wave equation for a single complex scalar \cite{Teukolsky:1973}.

Another useful result closely related to the algebraically special property is the separability of the geodesic and Klein-Gordon equations in the Kerr spacetime.

There are other interesting algebraically special spacetimes in 4d. For example, the C-metric, which describes a pair of black holes being accelerated apart by a conical deficit singularity (cosmic string) \cite{Kinnersley:1970zw}. Its Weyl tensor is type D, i.e. there are two pairs of coincident PNDs, just like the Kerr solution. It was this property that enabled the rotating generalization of the C-metric to be discovered \cite{Kinnersley:1969zza}. The C-metric is relevant also to higher-dimensional black holes: analytic continuation of a Kaluza-Klein generalization of the C-metric led to the discovery of black rings \cite{Emparan:2001wn}. Further examples of interesting algebraically special vacuum spacetimes include pp-waves and the Taub-NUT spacetime.  

Given the usefulness of the algebraically special property in 4d, it is natural to ask whether there is an analogous notion in $d>4$ dimensions.
 If so,  can it be exploited to solve the Einstein equation? Does it help in the study of perturbations of higher dimensional black holes?  

As we shall see, the algebraically special property can indeed be extended to $d>4$ dimensions. However, there are several inequivalent ways of doing this. This stems from the fact that there are several different methods for performing the Petrov classification of the Weyl tensor in 4d. These methods look very different but in fact are equivalent. However, when extended to $d>4$ dimensions they become inequivalent. 

A classification of the Weyl tensor in higher dimensions based on generalizing the notion of PNDs was introduced by Coley,  Milson, Pravda and Pravdova (CMPP) \cite{Coley:2004jv}. This is the classification which has received most attention so far. We shall review it in section \ref{sec:CMPP}. Section \ref{sec:desmet} reviews an alternative approach, due to De Smet \cite{DeSmet:2002fv}, based on generalizing the spinorial approach to the 4d classification. It turns out that this works only for the special case $d=5$. In general there does not seem to be any simple relation between the CMPP and De Smet approaches. In particular, the De Smet approach does not lead to any notion of preferred null directions. Nevertheless, Myers-Perry black holes \cite{Myers:1986un} are algebraically special with respect to both classifications. 
Other approaches to defining a notion of "algebraically special" in higher dimensions are discussed in section \ref{sec:other}.

The emphasis in this chapter will be on applications, particularly using the algebraically special property to construct new solutions of the Einstein equation, or to study perturbations of known solutions. We will consider only solutions of the {\it vacuum} Einstein equation, allowing for a cosmological constant.

\section{CMPP classification}

\label{sec:CMPP}

\subsection{The classification}

Coley, Milson, Pravda and Pravdova (CMPP) developed a scheme for classifying the Weyl tensor following closely the 4d approach based on principal null directions \cite{Coley:2004jv,Milson:2004jx} (see Ref. \cite{Coley:2007tp} for an earlier review). Although their approach is basis-independent, it is most easily described by introducing a null basis $e_0 \equiv \ell$, $e_1 \equiv n$, $e_i \equiv m_i$ ($i = 2, \ldots d-1$) where $\ell$ and $n$ are null vectors and $m_i$ are orthonormal spacelike vectors, satisfying the scalar products:
\be
 \ell^2 = n^2 = \ell \cdot m_i = n \cdot m_i = 0, \qquad \ell \cdot n = 1, \qquad m_i \cdot m_j = \delta_{ij}
\ee
Different null bases at a point are related by Lorentz transformations. It is convenient to give names to particular kinds of Lorentz transformation.  A {\it boost} is defined by 
\be
 \label{boost}
 \ell \rightarrow \lambda \ell, \qquad n \rightarrow \lambda^{-1} n, \qquad m_i \rightarrow m_i
\ee 
with $\lambda$ a scalar. A {\it spin} is a rotation of the spatial basis vectors: 
\be
\label{spin}
\ell \rightarrow \ell, \qquad n \rightarrow n, \qquad m_i \rightarrow X_{ij} m_j
\ee
 where $X_{ij}$ is an orthogonal matrix. A {\it null rotation about $\ell$} with parameters $z_i$ is defined by 
 \be
  \ell \rightarrow \ell, \qquad n \rightarrow n +z_i m_i - (1/2) z_i z_i \ell, \qquad m_i \rightarrow m_i - z_i \ell.
\ee 
 Together these three special types of transformation generate the full Lorentz group. 

A tensor component has {\it boost weight} $b$ if it scales as $\lambda^b$ under a boost. For example, consider the Weyl tensor components
\be
 C_{0i0j} = C_{\mu\nu\rho\sigma} \ell^\mu m_i^\nu \ell^\rho m_j^\sigma \rightarrow \lambda^2  C_{\mu\nu\rho\sigma} \ell^\mu m_i^\nu \ell^\rho m_j^\sigma = \lambda^2 C_{0i0j}
\ee
hence components $C_{0i0j}$ have $b=2$. In general, for a tensor with only "downstairs" indices, $b$ is the number of $0$ indices minus the number of $1$ indices. 

The symmetries of the Weyl tensor imply that it has components with $b$ ranging from $-2$ to $2$. We say that the vector field $\ell$ is a {\it Weyl aligned null direction} (WAND) iff the $b=2$ components of the Weyl tensor vanish. This definition is independent of how the other basis vectors are chosen. Indeed it is equivalent to the condition (\ref{pnd}). Hence, for $d=4$, a WAND is the same as a principal null direction.

One reason for employing different terminology when $d>4$ (i.e. "WAND" rather than principal null direction) is that WANDs behave rather differently when $d>4$. When $d=4$, there are 4 solutions to (\ref{pnd}) (although some may coincide). However, for $d>4$, this equation may admit no solutions, finitely many solutions, or even a continuous family of solutions. For example, the static Kaluza-Klein bubble (the product of a flat time direction with the Euclidean Schwarzschild solution) admits no WAND \cite{Godazgar:2009fi}. The Schwarzschild solution admits two WANDs. In the product spacetime $dS_3 \times S^2$, any null vector field tangent to $dS_3$ is a WAND so there is a continuous family \cite{Godazgar:2009fi}. 

In 4d, a spacetime is said to be algebraically special if it admits everywhere a {\it repeated} principal null direction (i.e. two or more null directions coincide). The higher dimensional analogue of this is a {\it multiple WAND}. We say that the null vector field $\ell$ is a multiple WAND if all $b=2,1$ Weyl components vanish. Once again, this is a basis-independent statement. We define a spacetime to be algebraically special if it admits (everywhere) a multiple WAND.\footnote{
Given that a spacetime might not admit even a WAND, CMPP defined a higher dimensional spacetime to be algebraically special if it admits a WAND (everywhere). However, more recent papers have not maintained this nomenclature for several reasons. First, it leads to tension with standard terminology in 4d. Second, there are examples of analytic spacetimes that admit a WAND in some open subset but not is some other open subset \cite{Pravda:2005kp,Godazgar:2009fi}. Third, the existence of a WAND is not sufficiently restrictive to be a useful tool to address the problems discussed in the introduction.
}
 
The algebraic type of the Weyl tensor is defined by the following table. 
\medskip

\begin{tabular}{l|l|l}
Algebraic type & Vanishing Weyl components & Comment \\
\hline
O & b=2,1,0,-1,-2 & Conformally flat\\
N & b=2,1,0,-1 & admits multiple WAND\\
III  & b=2,1,0 & admits multiple WAND\\
II & b=2,1& admits multiple WAND \\
I & b=2 & admits WAND\\
G & & admits no WAND\\
\end{tabular}
\medskip

For example, we say that the Weyl tensor is type III if one can choose $\ell$ so that all the $b=2,1,0$ components of the Weyl tensor vanish, but one can't choose $\ell$ so that all the $b=2,1,0,-1$ components vanish. It is type I if there is a WAND but not a multiple WAND. Types II, III, N and O admit a multiple WAND. This classification can be made at any point. The algebraic type of a {\it spacetime} is defined to be the least special type of the Weyl tensor at any point in the spacetime.

This  classification, based only on the vector $\ell$ is the so-called primary classification. One can now ask, given an $\ell$ defined by the primary classification, whether one can choose $n$ to be a WAND or a multiple WAND, corresponding respectively to the vanishing of the $b=-2$ or $b=-2,-1$ Weyl components. This leads to a more refined secondary classification. For our purposes, we shall be interested in secondary classification only in the case for which $\ell$ and $n$ both can be chosen to be multiple WANDs. In this case, the spacetime must be of primary type II (if not conformally flat). Such a spacetime is referred to as a type D spacetime. Note that the algebraic type of a spacetime as defined above agrees with the standard definition when $d=4$.

\subsection{Examples}

The 4d Schwarzschild and Kerr solutions are examples of type D spacetimes. Similarly, in $d>4$ dimensions, the Myers-Perry solution \cite{Myers:1986un} (and its generalization to include a cosmological constant \cite{Hawking:1998kw,Gibbons:2004uw} and NUT charge \cite{Chen:2006xh}) are type D \cite{Frolov:2003en,Hamamoto:2006zf}. Black rings are not algebraically special: Ref. \cite{Pravda:2005kp} found that the singly spinning black ring solution is type I in some region and type G in another region.\footnote{This emphasizes that the distinction between type I and type G probably is not useful (simpler examples of this behaviour were given in Ref. \cite{Godazgar:2009fi}). Given this example, one might wonder whether it is possible for a spacetime to be algebraically special in some open region and algebraically general in another open region. This seems unlikely: in simple examples, the condition for existence of a WAND reduces to an inequality involving the metric components. The type I/G behaviour just discussed corresponds to this inequality being satisfied in some region but not in another. However, the criterion for the existence of a {\it multiple} WAND is that certain combinations of metric components should vanish. If this happens in some open region, and the spacetime is analytic, then it must happen everywhere. It would be nice to have a proof of this in general.} 

Consider a warped product spacetime of the form
\be
 ds^2 = A(y)^2 g_{MN}(x) dx^M dx^N + B(x)^2 g_{AB}(y) dy^A dy^B
\ee
where $g_{MN}$ is Lorentzian and $g_{AB}$ is Riemannian. Ref \cite{Pravda:2007ty} showed that such a spacetime is type D (or O) if the Lorentzian factor is (i) two-dimensional; (ii) a three-dimensional Einstein spacetime; (iii) a type D Einstein spacetime. Hence a black $p$-brane obtained from the product of the Schwarzschild (or Myers-Perry) solution with $p$ flat dimensions must be type D. Similarly, product spacetimes such as $dS_p \times S^{d-p}$ ($p>1$) are type D.

Note that Wick rotation of a spacetime typically changes its algebraic type. For example, a 5d Schwarzschild black string can be Wick rotated to give a static Kaluza-Klein bubble spacetime (the product of the Euclidean Schwarzschild solution with a flat time direction). The former solution is type D but the latter is type G \cite{Godazgar:2009fi}. (An {\it expanding} KK bubble, given by Wick rotation of the 5d Schwarzschild spacetime \cite{Witten:1981gj} is a warped product with a 3d de Sitter factor and hence type D by the above theorem.)

\subsection{Goldberg-Sachs theorem}

Probably the most important result concerning 4d algebraically special solutions is the Goldberg-Sachs theorem:

\noindent {\it In a vacuum spacetime (allowing for a cosmological constant) that is not conformally flat, a null vector field is a repeated principal null direction if, and only if, it is geodesic and shearfree.}

Generalizations of this theorem exist, for example to include a Maxwell field. This theorem plays an important role in using the algebraically special property as a simplifying assumption to solve the Einstein equation. One introduces coordinates $(r,x^i)$ where $r$ is an affine parameter along the null geodesics whose existence is guaranteed by the theorem. Using the algebraically special property and the Einstein equation, one can determine the full dependence of the metric on $r$ \cite{Stephani:2003tm}. The Einstein equation then reduces to equations governing the dependence on the other 3 coordinates $x^i$.  This is a significant reduction in the difficulty of the problem although the reduced equations are still sufficiently complicated that one has to resort to extra assumptions to solve them.

The most spectacular use of the algebraically special property to solve the Einstein equation was Kinnersley's discovery of the most general type D solution \cite{Kinnersley:1969zza}. In this case, one has two principal null directions, both obeying the GS theorem and this is sufficiently restrictive that the Einstein equations can be solved explicitly. Kinnersley's result provides significant motivation for the study of higher-dimensional algebraically special solutions. If we want to emulate his work then the first step is to generalize the Goldberg-Sachs theorem.

The GS theorem as stated above does not extend to higher dimensions. There are simple examples of spacetimes which admit non-geodesic multiple WANDs (e.g. $dS_3 \times S^2$: {\it any} null vector field tangent to $dS_3$ is a multiple WAND irrespective of whether or not it is geodesic) and spacetimes which admit geodesic but shearing multiple WANDs (e.g. a 5d black string: the multiple WAND is a repeated PND of the 4d Schwarzschild spacetime; this expands isotropically in the Schwarzschild directions but not the string direction hence it is shearing). Nevertheless, it is clear that there {\it are} strong restrictions on the optical properties of a multiple WAND and understanding precisely what these are is the problem of formulating a higher-dimensional generalization of the GS theorem.

A first step towards such a generalization is to understand how the "geodesic part" of the theorem must be modified in higher dimensions. This has now been completed. For vacuum spacetimes of type III or N, it can be shown that the multiple WAND must be geodesic \cite{Pravda:2004ka}. For type II (or D), it need not be geodesic. However, in this case, it has been proved that there must exist another multiple WAND which {\it is} geodesic \cite{Durkee:2009nm}. In other words: a vacuum spacetime admits a multiple WAND if, and only if, it admits a geodesic multiple WAND. The $dS_3 \times S^2$ example is the prototype for this behaviour. This result implies that there is no loss of generality in restricting attention to geodesic multiple WANDs.

The next step is to understand the higher-dimensional analogue of the shearfree condition. This has not yet been achieved. However, there has been some partial progress. Recall that the expansion, rotation and shear are defined to be the trace, traceless symmetric part, and antisymmetric part of the matrix
\be
\label{rhodef}
 \rho_{ij} = \nabla_j \ell_i = m_i^\mu m_j^\nu \nabla_\nu \ell_\mu
\ee
The problem is to determine necessary (and, if possible, sufficient) conditions on $\rho_{ij}$ for $\ell$ to be a multiple WAND. So far, progress has been made only with the more special algebraic types. For a type N vacuum spacetime, it has been shown \cite{Pravda:2004ka} that one can choose the spatial basis vectors $m_i$ so that $\rho_{ij}$ is zero everywhere except for a $2\times 2$ block in the upper left corner of the form
\be
\label{2dblock}
 \left( \begin{tabular}{cc} $b$& $a$ \\ $-a$ & $b$ \end{tabular} \right)
\ee
Note that this $2 \times 2$ block has vanishing traceless symmetric part. However, the full $(d-2) \times (d-2)$ matrix $\rho_{ij}$ is not shearfree except when $d=4$. The same result applies for type III spacetimes when $d=5$ and also for $d>5$  subject to a certain "genericity" assumption \cite{Pravda:2004ka}.

Results have also been obtained for the case of a Kerr-Schild spacetime. Such a spacetime has a metric of the form
\be
 g_{\mu\nu} = \eta_{\mu\nu} + H k_\mu k_\nu
\ee
where $k_\mu$ is null with respect to the Minkowski metric $\eta_{\mu\nu}$ (which implies that it is null also with respect to $g_{\mu\nu}$), and $H$ is a function. This class of spacetimes includes Myers-Perry black holes. Vacuum Kerr-Schild spacetimes are algebraically special for any $d \ge 4$, with $k_\mu$ a geodesic multiple WAND \cite{Ortaggio:2008iq}. For these spacetimes, it has been shown that one can choose the basis vectors $m_i$ so that $\rho_{ij}$ is block diagonal, with $2\times 2$ blocks of the form (\ref{2dblock}) along the diagonal (the parameters $a,b$ can vary from block to block) and zeros elsewhere \cite{Ortaggio:2008iq,Malek:2010mh}.

These partial results are encouraging evidence that there does exist a simple canonical form for the matrix $\rho_{ij}$ defined by a multiple WAND. This will provide a higher-dimensional generalization of the "only if" part of the Goldberg-Sachs theorem. It is less clear at present whether the "if" part of the theorem also will generalize, in other words whether the canonical form for $\rho_{ij}$ forces $\ell$ to be a multiple WAND. Some partial results in this direction are known. For example, if $\ell$ is geodesic with vanishing rotation and shear then it must be a multiple WAND \cite{Podolsky:2006du,Ortaggio:2007eg}.

\subsection{Peeling theorem}

Some physical motivation for the study of algebraically special solutions in 4d comes from the peeling theorem. Consider a 4d asymptotically flat spacetime and a null geodesic that approaches future null infinity. Let $\ell$ denote the tangent to the geodesic. Then the Weyl tensor along this geodesic can be expanded in inverse powers of the affine parameter $r$ along the geodesic:
\be
 C_{\mu\nu\rho\sigma} = \frac{1}{r} C_{\mu\nu\rho\sigma}^{(1)} +  \frac{1}{r^2} C_{\mu\nu\rho\sigma}^{(2)} +  \frac{1}{r^3} C_{\mu\nu\rho\sigma}^{(3)} +  \frac{1}{r^4} C_{\mu\nu\rho\sigma}^{(4)} + {\cal O}\left( \frac{1}{r^5} \right)
\ee
where $C^{(1)}$ is type N, $C^{(2)}$ is type III, $C^{(3)}$ is type II (or D), in each case with repeated principal null direction $\ell$, and $C^{(4)}$ is type I, with principal null direction $\ell$. When first discovered, this suggested that the study of algebraically special spacetimes may be valuable for understanding the behaviour of spacetime far from a radiating source. 

An argument in Ref. \cite{Ortaggio:2009zt} suggests that this peeling property may not extend to higher dimensions. If an expansion of the above form does hold (with suitable powers of $r$) then the leading behaviour of $C_{\mu\nu\rho\sigma}$ near null infinity is type N. Therefore one would expect that, at large $r$, $\rho_{ij}$ should have the form appropriate to a type N solution. As discussed above, this is a rank 2 matrix, i.e., it singles out a 2d subspace. However, in the far-field of a general radiating spacetime, one might not expect such a preferred subspace to occur. If this is correct then the leading behaviour cannot be type N. A similar argument excludes type III (at least for $d=5$).\footnote{Alternatively, type N or type III might occur but with $\rho_{ij}=0$.} Of course, this argument does not apply to asymptotically {\it locally} flat spacetimes, e.g. Kaluza-Klein asymptotics, for which one would expect $\rho_{ij}$ to be degenerate near infinity. It seems likely that a peeling theorem would exist for spacetimes with boundary conditions appropriate to a KK reduction to 4d.

\subsection{Finding new solutions}

One of the main motivations for the study of algebraically special solutions in $d>4$ dimensions is the hope that the algebraically special property will make the Einstein equation easier to solve, leading to the discovery of interesting new explicit solutions. In contrast with many other methods for solving the Einstein equation, this approach can cope with a non-zero cosmological constant. So far, there have been just a few investigations in this direction. 

In 4d, the Goldberg-Sachs theorem implies that a repeated PND $\ell$ is geodesic and shearfree. In the simplest classes of algebraically special solution, the rotation of $\ell$ also vanishes. This implies that $\ell$ is orthogonal to a family of null hypersurfaces $u={\rm constant}$. One can introduce coordinates $(r,y^i)$ on these surfaces where $r$ is an affine parameter along the geodesics with tangent $\ell$ and $i=1,2$. This gives a coordinate chart $(u,r,y^i)$ on spacetime. Since the shear and rotation vanish, $\rho_{ij} = (1/2) \rho_{kk} \delta_{ij}$ where $\rho_{kk}$ is the expansion. These spacetimes fall into two classes \cite{Stephani:2003tm}: Robinson-Trautman spacetimes, for which $\rho_{kk} \ne 0$, and Kundt spacetimes for which $\rho_{kk}=0$ (and hence $\rho_{ij}=0$).  In both classes, the vacuum Einstein equation determines fully the $r$-dependence of the metric, and reduces to nonlinear PDEs for a small number of functions of $(u,y^i)$. The Robinson-Trautman family contains the Schwarzschild solution and the C-metric and certain time-dependent generalizations of these spacetimes. The Kundt family describes various generalized gravitational wave solutions.

In $d>4$ dimensions, some general properties of vacuum solutions with a hypersurface orthogonal (geodesic) multiple WAND were investigated in Ref. \cite{Pravdova:2008gp}. The $r$-dependence of such solutions was fully determined. As we discussed above, it is no longer true that the shear of the multiple WAND must vanish in general. However, if one makes the assumption that the shear {\it does} vanish then one can obtain higher-dimensional generalizations of the Robinson-Trautman and Kundt families of solutions. 

The equations governing $d>4$ dimensional vacuum Kundt solutions were obtained in Ref. \cite{Podolsky:2008ec}. The Robinson-Trautman case was investigated in Ref. \cite{Podolsky:2006du}. Perhaps disappointingly, the latter reference discovered that this class of solutions is considerably less rich in $d>4$ dimensions; it contains the higher-dimensional Schwarzschild solution but no time-dependent generalizations. Furthermore, there is nothing that could be identified as a higher-dimensional C-metric. However, this lack of richness probably occurs because the shear-free condition is too restrictive in higher dimensions: we know that a multiple WAND need not be shear-free. 

A promising avenue for future research would be to determine the canonical form for $\rho_{ij}$ for the special case of a hypersurface-orthogonal multiple WAND, which is probably much easier than finding a full higher-dimensional generalization of the Goldberg-Sachs theorem. This could then be used as the starting point for integrating the Einstein equation using the above method.

A different approach is to combine the algebraically special property with the assumption that the metric admits certain symmetries. Ref. \cite{Godazgar:2009fi} investigated the class of {\it axisymmetric} metrics, defined as those admitting a $SO(d-2)$ isometry group with $S^{d-3}$ orbits. This includes the interesting class of {\it static} axisymmetric solutions, with $R\times SO(d-2)$ isometry group. Such solutions would include the solution describing a brane-world black hole in the Randall-Sundrum model, which is thought to be related to a higher-dimensional C-metric. Note that the general static axisymmetric solution of the Einstein equation is not known (except for $d=4$ with vanishing cosmological constant). However Ref. \cite{Godazgar:2009fi} showed that if one demands that the metric be algebraically special, as well as axisymmetric (not necessarily static), then the Einstein equation (with cosmological constant) can be solved. Most of the solutions discovered were already known (e.g. generalizations of the Schwarzschild black hole, black string or expanding KK bubble) However, there is one class (corresponding to axisymmetric Kundt solutions) which is less familiar. It is specified by solutions of certain ODEs that cannot be solved analytically. Some spacetimes in this class have AdS/CFT applications, e.g., the gravitational dual of a CFT (vacuum state) in $R \times S^1 \times S^2$ \cite{Copsey:2006br} or $AdS_2 \times S^2$ \cite{Kaus:2009cg}.

\subsection{Perturbations}

As discussed in the introduction, a major motivation for trying to exploit the algebraically special property in higher dimensions is the application to the study of linearized gravitational perturbations of black hole spacetimes. In 4d, the Weyl tensor is encoded in the Newman-Penrose scalars $\Psi_A$, $A=0,1,2,3,4$. The algebraically special property has two important consequences for perturbations \cite{Teukolsky:1973}. 

First, choosing the null basis so that $\ell$ is the repeated PND of the unperturbed spacetime, $\delta \Psi_0$ (the perturbation of $\Psi_0$) is gauge invariant under both infinitesimal diffeomorphisms and infinitesimal basis transformations. Note that this quantity is a complex scalar and therefore has two degrees of freedom, the same number as the gravitational field in 4d. It is therefore plausible that it captures all information about generic metric perturbations.  This is indeed the case: it can be shown that knowledge of the perturbation in $\Psi_0$ is sufficient to reconstruct the metric perturbation up to the freedom to add a perturbation for which $\delta \Psi_0$ vanishes \cite{Wald:1978vm}. In the Kerr spacetime, it can be shown \cite{waldtypeDpert} that the only perturbations for which $\delta\Psi_0=0$ and are regular at the future horizon and decay at infinity are the modes corresponding to infinitesimal changes in the mass and angular momentum of the black hole. 

Second,  $\delta \Psi_0$ can be {\it decoupled} from the other $\delta \Psi_A$. $\delta \Psi_0$ satisfies a second order, linear, homogeneous wave equation in the unperturbed spacetime. Remarkably, if one Fourier analyzes this equation, i.e., assumes dependence proportional to $\exp(-i\omega t + i m \phi)$ then it can be separated, and reduced to ODEs governing the $r$ and $\theta$ dependence.

How much of this extends to perturbations of algebraically special vacuum solutions in $d>4$ dimensions? 
Choose a null basis in which $\ell$ is a multiple WAND and let $\Omega_{ij} = C_{0i0j}$. This is a traceless symmetric matrix. In 4d it is equivalent to $\Psi_0$. Ref. \cite{Durkee:2010qu} showed that the gauge-invariance properties of $\delta \Psi_0$ are satisfied also by $\delta \Omega_{ij}$. Note that this quantity has the same number of degrees of freedom as the linearized gravitational field. The 4d case suggests that  knowledge of $\delta \Omega_{ij}$ will be sufficient to determine the metric perturbation, up to the freedom to add a small number of "non-generic" modes. This remains to be shown. In any case, the fact that $\delta \Omega_{ij}$ is local and gauge invariant should make it useful in any study of perturbations of Myers-Perry black holes. 

Ref. \cite{Durkee:2010qu} analyzed also the conditions under which $\delta \Omega_{ij}$ satisfies a decoupled equation of motion. The result is that the conditions for decoupling in $d>4$ dimensions are considerably stronger than in 4d. In 4d, one finds only that the multiple PND should be geodesic and shearfree, which is guaranteed by the Goldberg-Sachs theorem. But for $d>4$, one finds that the multiple WAND should be geodesic with vanishing expansion, rotation and shear, i.e., $\rho_{ij}=0$. In other words, decoupling of $\delta \Omega_{ij}$ occurs if, and only if, the spacetime is a Kundt spacetime. Unfortunately, black hole spacetimes are not Kundt spacetimes so decoupling does not occur even for the Schwarzschild spacetime in $d>4$ dimensions.\footnote{
This is not in contradiction with the results of Ref. \cite{Kodama:2003jz} for Schwarzschild perturbations because Ref. \cite{Kodama:2003jz} did not use the local quantity $\delta \Omega_{ij}$ but instead used a non-local scalar/vector/tensor decomposition exploiting spherical symmetry.}

Although black holes are not Kundt spacetimes, the {\it near-horizon geometry} of an extreme black hole {\it is} a Kundt spacetime \cite{Durkee:2010qu}. Therefore one can use the decoupled equation for $\delta \Omega_{ij}$ to study perturbations of near-horizon geometries. This was explored in Ref. \cite{Durkee:2010ea}.

It is conceivable that there is some other gauge invariant quantity that does decouple, and reduces to $\delta \Psi_0$ when $d=4$. That this might be possible is suggested by results for the Schwarzschild spacetime for which there exist additional local gauge-invariant combinations of Weyl components and connection components (e.g. $\rho_{ij}$) \cite{Durkee:2010qu}. These combinations vanish identically for $d=4$. Perhaps it is possible to add such a combination to $\delta \Omega_{ij}$ to obtain a quantity that does satisfy a decoupled equation. 

\subsection{NP and GHP Formalisms}

In 4d, the Newman-Penrose (NP) formalism \cite{Newman:1961qr} is a convenient framework for performing certain calculations in GR, particularly those involving preferred null directions. Therefore it plays a central role in the study of algebraically special solutions.
The idea is simple: the components of any tensor with respect to a null basis are scalars (since they are defined by contraction of the tensor with the basis vectors). Hence by contracting with basis vectors one can reduce tensorial equations, involving covariant derivatives, to scalar equations involving only partial derivatives. A higher-dimensional generalization of the NP formalism has been developed in Refs. \cite{Pravda:2004ka,Ortaggio:2007eg,Coley:2004hu}. For the special case of $d=5$, a spinor based approach was presented in \cite{GomezLobo:2009ct}.

Often, and especially when studying algebraically special spacetimes, there exist one or more preferred null directions but no preferred spatial directions. The NP formalism has the drawback that it requires a specific choice of spatial basis vectors and does not maintain covariance with respect to spins (i.e. rotations of the spatial basis vectors). Furthermore, it does not maintain covariance with respect to boosts (which rescale the null basis vectors). These deficiencies were remedied in an improved formalism introduced by Geroch, Held and Penrose (GHP) \cite{Geroch:1973am}. A higher-dimensional generalization of this formalism was introduced in Ref. \cite{Durkee:2010xq}.

In the GHP approach, we say that an object $T_{i_1 \ldots i_s}$ is a {\it GHP scalar} if it transforms with a definite boost weight $b$ under a boost (\ref{boost}) and transforms under a spin (\ref{spin}) as
\be
 T_{i_1 \ldots i_s } \rightarrow X_{i_1 j_1} \ldots X_{i_s j_s} T_{j_1 \ldots j_s}.
\ee
We say that such a quantity has {\it spin} $s$. For example, the quantity $\rho_{ij}$ defined by (\ref{rhodef}), encoding the expansion, rotation and shear of $\ell$, is a GHP scalar with $b=1$ and $s=2$. Another important GHP scalar is
\be
 \kappa_i = \nabla_0 \ell_i = m_i^\mu \ell^\nu \nabla_\nu \ell_\mu
\ee
which has $b=2$ and $s=1$. This measures the failure of $\ell$ to be geodesic: $\ell$ is geodesic if, and only if, $\kappa_i = 0$. Not all NP scalars are GHP scalars. For example the NP scalar $\nabla_0 \ell_1 = n^\mu \ell^\nu \nabla_\nu \ell_\mu$ transforms inhomogeneously under a boost and therefore is not a GHP scalar.

One can introduce derivative operators which map GHP scalars to GHP scalars, and write out components of e.g. the Bianchi identity using these derivatives  \cite{Durkee:2010xq}. The use of this formalism considerably simplifies the analysis of algebraically special solutions, e.g., their perturbations  \cite{Durkee:2010qu}.

\section{De Smet classification}

\label{sec:desmet}

\subsection{The classification}

In 4d, the most straightforward explanation of the Petrov classification makes use of the 2-component spinor formalism.\footnote{We shall use $A,B, \ldots$ to denote spinor indices for general $d$. A spinor is denoted with a superscript index: $\epsilon^A$, dual spinors have subscript indices. The gamma matrices are denoted $(\Gamma^a)^A{}_B$ where $a,b,c,\ldots$ refer to an orthonormal basis. Spinor indices are lowered with $C_{AB}$ (the charge conjugation matrix) and raised with $C^{-1}$.} In this approach, the Weyl tensor is equivalent to the Weyl spinor $\Psi_{ABCD}$, which is totally symmetric in spinor indices. Therefore specifying the Weyl tensor at a point is equivalent to specifying a homogeneous quartic polynomial in two complex variables $(w,z)$:
\be
\label{poly}
 P = \Psi_{ABCD} \epsilon^A \epsilon^B \epsilon^C \epsilon^D
\ee
where $\epsilon^A = (w,z)$. By the fundamental theorem of algebra, this polynomial can be factorized (e.g. first divide through by $w^4$ to get a polynomial in $z/w$) and hence there exist (dual) spinors $\kappa_1, \ldots,\kappa_4$ such that
\be
 \Psi_{ABCD} = (\kappa_1)_{(A}  (\kappa_2)_{B} (\kappa_3)_{C} (\kappa_4)_{D)}. 
\ee
In 4d, a 2-component spinor defines a null vector via $\ell^a = \bar{\kappa} \Gamma^a \kappa$.
Hence $\kappa_1, \ldots \kappa_4$ define 4 null vectors. These are the principal null directions. The Petrov classification is equivalent to classifying the polynomial $P$ by the multiplicity of its factors, e.g., a type N spacetime has a single factor of multiplicity 4 whereas a type D spacetime has two factors of multiplicity 2.

The De Smet classification is a 5d generalization of this spinorial approach \cite{DeSmet:2002fv}. However, in contrast with the 4d case, the spinorial classification is not equivalent to the classification based on null directions. 

In 5d, one must work with Dirac spinors, which have 4 complex components. One can choose a representation in which $C$ and $C\Gamma^a$ are antisymmetric, and $C\Gamma^{ab}= C\Gamma^{[a} \Gamma^{b]}$ are symmetric. From the Weyl tensor one can define a Weyl spinor
\be
 \Psi_{ABCD} = (C\Gamma^{ab})_{AB} (C\Gamma^{cd})_{CD} C_{abcd}
\ee 
This is manifestly symmetric in $AB$ and in $CD$ and under interchange of $AB$ with $CD$. In fact it is totally symmetric. De Smet argued this using  an "accidental" isomorphism of $Spin(1,4)$ with a symplectic group. A more direct proof exploits the Fierz identity \cite{Godazgar:2010ks}. Either method works only for $d=5$.

Just as in 4d, the symmetry of $\Psi_{ABCD}$ implies that it is equivalent to a homogeneous quartic polynomial defined by (\ref{poly}). But there is an important difference with the 4d case: $\epsilon^A$ now has 4 complex components so $P$ is a polynomial in 4 complex variables. No longer can the fundamental theorem of algebra be invoked to reduce it to a product of linear factors. Instead, De Smet proposed that one should attempt to factorize $P$ into lower degree polynomials, and classify the Weyl tensor according to the degree and multiplicity of the factors. For example, if $P$ factorizes into the product of two distinct quadratic polynomials, the Weyl tensor is said to be of type ${\bf 22}$. An underline denotes a repeated factor so ${\bf \underline{22}}$ implies that $P$ is the square of a quadratic polynomial. If $P$ does not factorize then the Weyl tensor is algebraically general, denoted ${\bf 4}$. The algebraic type of a {\it spacetime} is defined to be the least special type of the Weyl tensor in the spacetime.

Various vacuum spacetimes have been classified according to the De Smet scheme. The Schwarzschild \cite{DeSmet:2002fv} and Myers-Perry  \cite{DeSmet:2003kt} solutions are type ${\bf \underline{22}}$. The Schwarzschild black string is type ${\bf 22}$ \cite{DeSmet:2002fv}. Black rings are algebraically general (type {\bf 4}) \cite{Godazgar:2010ks}. 

The Weyl tensor has 35 independent real components whereas a symmetric spinor $\Psi_{ABCD}$ has 35 complex components. Hence the equivalence of these objects implies that $\Psi_{ABCD}$ must satisfy a reality condition worked out in Ref. \cite{Godazgar:2010ks}. This implies restrictions on the possible factorizations of $P$. For example ${\bf \underline{11} 11}$ is not consistent with the reality condition.\footnote{\label{reality} A similar reality condition arises in 4d in Riemannian signature. In general one can write a 4d Weyl tensor in terms of $\Psi_{ABCD}$ and $\bar{\Psi}_{A'B'C'D'}$. In Lorentzian signature, reality of the Weyl tensor implies that these objects are related by complex conjugation. In Riemannian signature, they are independent but must each obey a reality condition which restricts their types to be $I$, $D$ or $O$ \cite{hacyan,Karlhede:1985ds}.} 
Incidentally, this counting of degrees of freedom shows that only for sufficiently small $d$ can the Weyl tensor be equivalent to a valence 4 symmetric spinor, simply because the number of independent components of the latter grows much more rapidly with $d$ than the number of components of the Weyl tensor.

Ref \cite{Godazgar:2010ks} discussed relations between the CMPP and De Smet classifications. It was found that a Weyl tensor belonging to a given CMPP class must belong to a restricted set of possible De Smet classes and vice versa.

\subsection{Finding new solutions}

 De Smet's motivation for introducing his classification was that the resulting algebraically special property might assist in solving the Einstein equation. As discussed above, the general static, axisymmetric, vacuum solution is not known for $d>4$. However, De Smet showed that if one imposes the additional condition that the spacetime is type ${\bf 22}$ or ${\bf \underline{22}}$ then the general solution can be found \cite{DeSmet:2002fv}, including the case with a cosmological constant \cite{DeSmet:2003bt}. Unfortunately, the list of solutions that results does not contain anything new. 

A different approach is to try to determine all vacuum solutions of a given De Smet type. This has been done for the simplest non-trivial type consistent with the reality condition, which is ${\bf \underline{11} \, \underline{11}}$ \cite{Godazgar:2010ks}. In this case, the Weyl spinor can be written in terms of a single spinor $\epsilon$ as $\Psi_{ABCD} = \epsilon_{(A} \epsilon_B \bar{\epsilon}_C \bar{\epsilon}_{D)}$. The existence of a globally defined spinor field makes the analysis reminiscent of the analysis of supersymmetric solutions of 5d supergravity performed in Ref. \cite{Gauntlett:2002nw}. One defines a real scalar $f = \bar{\epsilon} \epsilon$, a real vector $V_a = i \bar{\epsilon} \Gamma_a \epsilon$ and a real 2-form  $F_{ab} =i \bar{\epsilon} \Gamma_{ab} \epsilon$. The Fierz identity implies various algebraic relations between these objects, for example that $V^2 = -f^2$, so $V$ must be timelike or null. The Weyl tensor can be written as an expression quadratic in $f$, $V$ and $F$. The Bianchi identity then imposes differential conditions on these objects. The idea is to exploit these conditions to help solve the Einstein equation (with $\Lambda$).

In the case in which $V$ is globally null, the solutions must be Kundt solutions that are type N in the CMPP classification \cite{Godazgar:2010ks}. The more interesting case is when $V$ is timelike. In this case, the solutions are of the form \cite{Godazgar:2010ks} $ds^2 = -dt^2 + A(t)^2 ds_4^2$ where $ds_4^2$ is a 4d Einstein space of Petrov type\footnote{
Recall footnote \ref{reality}: for a Riemannian manifold, the Petrov types of $\Psi$ and $\bar{\Psi}$ are independent.} $(D,O)$ and $A(t)$ is some simple function of $t$. This class includes the Kaluza-Klein monopole spacetime \cite{Gross:1983hb,Sorkin:1983ns}.

This example illustrates how spacetimes that are algebraically special in the De Smet scheme admit various globally defined tensor fields in spacetimes. The types of field that arise are different for each De Smet class. Another example is type ${\bf \underline{22} }$, for which the Weyl tensor can be written entirely in terms of a certain 2-form \cite{Godazgar:2010ks}. The Bianchi identity imposes differential conditions on this 2-form. It will be interesting to see whether these kinds of condition are restrictive enough to enable one to determine the most general solution of the Einstein equation in various De Smet classes.

\section{Other approaches}

\label{sec:other}

\subsection{Bivector classification}

The symmetries of the Weyl tensor implies that it can be used to map 2-forms (bivectors) to 2-forms:
\be
 \omega_{\mu\nu} \rightarrow \omega'_{\mu\nu} =  \frac{1}{2} C_{\mu\nu}{}^{\rho\sigma} \omega_{\rho\sigma} 
\ee
If we introduce a basis for 2-forms and denote basis indices with capital letters then we can write this as
\be
 \omega'_A = C_A{}^B \omega_B
\ee
In this approach, the idea is to classify the Weyl tensor by bringing the matrix $C_A{}^B$ to a canonical form that depends on (but is not fully determined by) its eigenvalues. This was how Petrov performed the classification of the Weyl tensor in 4d. 

Use of this method in higher dimensions was discussed in Ref. \cite{Coley:2009is}. The resulting classification that appears to be distinct from the CMPP (and De Smet) classifications. 

\subsection{Hidden symmetries}

The subject of hidden symmetries of higher-dimensional black holes has received a lot of attention recently, see Ref. \cite{Krtous:2008tb} and references therein. In this section we shall discuss only how this topic relates to algebraic classification of the Weyl tensor.

We start with a definition: a {\it conformal Killing-Yano 2-form} is a 2-form $\phi$ satisfying
\be
\nabla_\mu \phi_{\nu\rho} = \tau_{\mu\nu\rho} +\frac{2}{d-1} g_{\mu[\nu} K_{\rho]}
\ee
for some 3-form $\tau$ and 1-form $K$. If $K=0$ then $\phi$ is a {\it Killing-Yano 2-form}. The remarkable fact that the geodesic equation, Klein-Gordon equation and Dirac equation are all separable in the Kerr spacetime is a consequence of the existence of a Killing-Yano 2-form in this spacetime. 

Other 4d type D spacetimes, e.g., the C-metric admit a conformal Killing-Yano 2-form (which leads to separation of variables for {\it null} geodesics). In fact, in 4d, the type D property is equivalent to the existence of a {\it non-degenerate} conformal Killing-Yano 2-form. This led Ref. \cite{Mason:2008ih} to propose that one could take existence of such a 2-form as the higher-dimensional generalization of the type D property. It was shown (subject to an assumption that the eigevalues of $\phi^\mu{}_\nu$ are distinct) that this definition implies that the Weyl tensor must be type D in the CMPP classification. I am not aware of any results in the opposite direction.

There has been some progress in using the type D property as defined by Ref. \cite{Mason:2008ih} to solve the Einstein equation. This starts from the additional assumption that $\phi$ should be closed, i.e. $\tau=0$ above. ($\phi$ is then called a {\it principal} conformal Killing-Yano 2-form.) It has been shown \cite{Houri:2007xz,Krtous:2008tb} that the most general solution of the vacuum Einstein equation that admits such a 2-form is the Myers-Perry solution generalized to include NUT charge and cosmological constant \cite{Chen:2006xh}. Of course this was a solution which was already known. It would be interesting to know whether further progress could be made in solving the Einstein equation in the general case for which $\tau \ne 0$.

\subsection{Optical structures}

The motivation here comes from the Goldberg-Sachs theorem: the starting point is to invent a higher-dimensional generalization of the 4d condition that the spacetime should admit a shear-free null geodesic congruence. We shall refer to such a generalization as an optical structure. Given the existence of such a structure, one can ask what restrictions this imposes on the Weyl tensor.

This approach was adopted in Ref. \cite{Hughston:1988nz} (see also Ref. \cite{trautman}), which proposed that an appropriate structure in $d=2m$ dimensions is a $m$-dimensional distribution ${\cal D}$ of the complexified tangent bundle of spacetime, that is integrable ($[{\cal D}, {\cal D}] \subset {\cal D}$) and totally null ($g(V,W)=0$ for any $V,W \in {\cal D}$). The intersection ${\cal D} \cap \bar{\cal D}$ is then 1-dimensional and consists of real null vectors. It can be shown that these are geodesic. In 4d, this definition is equivalent to the existence of a shear-free null geodesic congruence. Definitions appropriate for odd $d$ were given in Refs. \cite{Mason:2008ih,TaghaviChabert:2010bm}. 

Optical structures are relevant for higher-dimensional black holes. Ref. \cite{Mason:2008ih} showed that the Myers-Perry-(AdS) solution admits such structures. Indeed, they are closely related to the conformal Killing-Yano tensors that such solutions possess. Black rings also possess an optical structure \cite{TaghaviChabert:2010bm}. 

The existence of an optical structure imposes certain restrictions on the Weyl tensor \cite{Mason:2008ih,TaghaviChabert:2010bm}. In order to have a generalization of the Goldberg-Sachs theorem, one might seek necessary and sufficient conditions on the Weyl tensor for the existence of an optical structure. Progress was made for $d=5$ in Ref. \cite{TaghaviChabert:2010bm}, which presented conditions on the Weyl tensor sufficient\footnote{
Subject to a "genericity" assumption.} for the existence of an optical structure. It was proposed that these conditions should be adopted as the definition of "algebraically special" for $d=5$ and possibly $d>5$. (In 5d, this definition is stronger than the CMPP definition \cite{TaghaviChabert:2010bm}.) However, it was observed that black rings are a counterexample to the converse of this result: they admit an optical structure but are not algebraically special.

\section{Outlook}

Various approximate techniques point to the existence of large families of new black hole solutions in higher dimensions. If we are to go beyond perturbative methods to construct these solutions then we must either resort to numerics or develop new techniques for solving the Einstein equation analytically. Results in 4d suggest that exploiting the algebraically special property to simplify the Einstein equation is the most promising technique available for finding new analytic solutions. As discussed above, there are several different notions of "algebraically special" in higher dimensions and therefore, possibly, several different ways of simplifying the Einstein equation.
It will be interesting to see what new solutions are discovered by these methods.

\medskip

{\bf Acknowledgments}

I am grateful to Mahdi Godazgar, Vojtech Pravda and Alena Pravdova  for comments on a draft of this chapter.


\begin{thebibliography}{99}

\bibitem{Kerr:1963ud}
  R.~P.~Kerr,
  Phys.\ Rev.\ Lett.\  {\bf 11}, 237 (1963).

\bibitem{Teukolsky:1973}
S.~A. Teukolsky,
\newblock Astrophys. J. {\bf 185}, 635 (1973).

\bibitem{Kinnersley:1970zw}
  W.~Kinnersley and M.~Walker,
  Phys.\ Rev.\  D {\bf 2}, 1359 (1970).

\bibitem{Kinnersley:1969zza}
  W.~Kinnersley,
  J.\ Math.\ Phys.\  {\bf 10}  1195 (1969).

\bibitem{Emparan:2001wn}
  R.~Emparan, H.~S.~Reall,
  Phys.\ Rev.\ Lett.\  {\bf 88}, 101101 (2002).
  [hep-th/0110260].

\bibitem{Coley:2004jv}
  A.~Coley, R.~Milson, V.~Pravda and A.~Pravdova,
  Class.\ Quant.\ Grav.\  {\bf 21}, L35 (2004)
  [arXiv:gr-qc/0401008].

\bibitem{DeSmet:2002fv}
  P.~-J.~De Smet,
  Class.\ Quant.\ Grav.\  {\bf 19}, 4877-4896 (2002).
  [hep-th/0206106].

\bibitem{Myers:1986un}
  R.~C.~Myers and M.~J.~Perry,
  Annals Phys.\  {\bf 172}, 304 (1986).

\bibitem{Milson:2004jx}
  R.~Milson, A.~Coley, V.~Pravda and A.~Pravdova,
  Int.\ J.\ Geom.\ Meth.\ Mod.\ Phys.\  {\bf 2}, 41 (2005)
  [arXiv:gr-qc/0401010].

\bibitem{Coley:2007tp}
  A.~Coley,
  Class.\ Quant.\ Grav.\  {\bf 25}, 033001 (2008)
  [arXiv:0710.1598 [gr-qc]].

\bibitem{Godazgar:2009fi}
  M.~Godazgar and H.~S.~Reall,
  Class.\ Quant.\ Grav.\  {\bf 26}, 165009 (2009)
  [arXiv:0904.4368 [gr-qc]].

\bibitem{Pravda:2005kp}
  V.~Pravda and A.~Pravdova,
  Gen.\ Rel.\ Grav.\  {\bf 37}, 1277 (2005)
  [arXiv:gr-qc/0501003].

\bibitem{Hawking:1998kw}
  S.~W.~Hawking, C.~J.~Hunter and M.~Taylor,
  Phys.\ Rev.\  D {\bf 59}, 064005 (1999)
  [arXiv:hep-th/9811056].

\bibitem{Gibbons:2004uw}
  G.~W.~Gibbons, H.~Lu, D.~N.~Page and C.~N.~Pope,
  J.\ Geom.\ Phys.\  {\bf 53}, 49 (2005)
  [arXiv:hep-th/0404008].

\bibitem{Chen:2006xh}
  W.~Chen, H.~Lu and C.~N.~Pope,
  Class.\ Quant.\ Grav.\  {\bf 23}, 5323 (2006)
  [arXiv:hep-th/0604125].

\bibitem{Frolov:2003en}
  V.~P.~Frolov, D.~Stojkovic,
  Phys.\ Rev.\  {\bf D68}, 064011 (2003).
  [gr-qc/0301016].

\bibitem{Hamamoto:2006zf}
  N.~Hamamoto, T.~Houri, T.~Oota, Y.~Yasui,
  J.\ Phys.\ A {\bf A40}, F177-F184 (2007).
  [hep-th/0611285].

\bibitem{Pravda:2007ty}
  V.~Pravda, A.~Pravdova and M.~Ortaggio,
  Class.\ Quant.\ Grav.\  {\bf 24}, 4407 (2007)
  [arXiv:0704.0435 [gr-qc]].

\bibitem{Witten:1981gj}
  E.~Witten,
  Nucl.\ Phys.\  {\bf B195}, 481 (1982).

\bibitem{Stephani:2003tm}
  H.~Stephani, D.~Kramer, M.~A.~H.~MacCallum, C.~Hoenselaers, E.~Herlt,
  Cambridge, UK: Univ. Pr. (2003) 701 P.

\bibitem{Pravda:2004ka}
  V.~Pravda, A.~Pravdova, A.~Coley and R.~Milson,
  Class.\ Quant.\ Grav.\  {\bf 21}, 2873 (2004)
  [Erratum-ibid.\  {\bf 24}, 1691 (2007)]
  [arXiv:gr-qc/0401013].

\bibitem{Durkee:2009nm}
  M.~Durkee and H.~S.~Reall,
  Class.\ Quant.\ Grav.\  {\bf 26}, 245005 (2009)
  [arXiv:0908.2771 [gr-qc]].

\bibitem{Ortaggio:2008iq}
  M.~Ortaggio, V.~Pravda and A.~Pravdova,
  Class.\ Quant.\ Grav.\  {\bf 26}, 025008 (2009)
  [arXiv:0808.2165 [gr-qc]].

\bibitem{Malek:2010mh}
  T.~Malek and V.~Pravda,
  Class.\ Quant.\ Grav.\  {\bf 28}, 125011 (2011)
  [arXiv:1009.1727 []].


\bibitem{Podolsky:2006du}
  J.~Podolsky, M.~Ortaggio,
  Class.\ Quant.\ Grav.\  {\bf 23}, 5785-5797 (2006).
  [gr-qc/0605136].


\bibitem{Ortaggio:2009zt}
  M.~Ortaggio, V.~Pravda and A.~Pravdova,
  Phys.\ Rev.\  D {\bf 80}, 084041 (2009)
  [arXiv:0907.1780 [gr-qc]].

\bibitem{Pravdova:2008gp}
  A.~Pravdova, V.~Pravda,
  Class.\ Quant.\ Grav.\  {\bf 25}, 235008 (2008).
  [arXiv:0806.2423 [gr-qc]].

\bibitem{Podolsky:2008ec}
  J.~Podolsky and M.~Zofka,
  Class.\ Quant.\ Grav.\  {\bf 26}, 105008 (2009)
  [arXiv:0812.4928 [gr-qc]].

\bibitem{Copsey:2006br}
  K.~Copsey and G.~T.~Horowitz,
  JHEP {\bf 0606}, 021 (2006)
  [arXiv:hep-th/0602003].

\bibitem{Kaus:2009cg}
  A.~Kaus and H.~S.~Reall,
  JHEP {\bf 0905}, 032 (2009)
  [arXiv:0901.4236].

\bibitem{Wald:1978vm}
  R.~M.~Wald,
  Phys.\ Rev.\ Lett.\  {\bf 41}, 203 (1978).

\bibitem{waldtypeDpert}
R.~M.~Wald, 
J. Math. Phys.
{\bf 14} (1973) 1453.

\bibitem{Durkee:2010qu}
  M.~Durkee and H.~S.~Reall,
  Class.\ Quant.\ Grav.\  {\bf 28}, 035011 (2011)
  [arXiv:1009.0015 [gr-qc]].

\bibitem{Kodama:2003jz}
  H.~Kodama and A.~Ishibashi,
  Prog.\ Theor.\ Phys.\  {\bf 110}, 701 (2003)
  [arXiv:hep-th/0305147].

\bibitem{Durkee:2010ea}
  M.~Durkee and H.~S.~Reall,
  [arXiv:1012.4805 [hep-th]].

\bibitem{Newman:1961qr}
  E.~Newman and R.~Penrose,
  J.\ Math.\ Phys.\  {\bf 3}, 566 (1962).

\bibitem{Ortaggio:2007eg}
  M.~Ortaggio, V.~Pravda and A.~Pravdova,
  Class.\ Quant.\ Grav.\  {\bf 24}, 1657 (2007)
  [arXiv:gr-qc/0701150].

\bibitem{Coley:2004hu}
  A.~Coley, R.~Milson, V.~Pravda and A.~Pravdova,
  Class.\ Quant.\ Grav.\  {\bf 21}, 5519 (2004)
  [arXiv:gr-qc/0410070].

\bibitem{GomezLobo:2009ct}
  A.~P.~Gomez-Lobo and J.~M.~Martin-Garcia,
  J.\ Math.\ Phys.\  {\bf 50}, 122504 (2009)
  [arXiv:0905.2846 [gr-qc]].

\bibitem{Geroch:1973am}
  R.~P.~Geroch, A.~Held and R.~Penrose,
  J.\ Math.\ Phys.\  {\bf 14}, 874 (1973).

\bibitem{Durkee:2010xq}
  M.~Durkee, V.~Pravda, A.~Pravdova and H.~S.~Reall,
  Class.\ Quant.\ Grav.\  {\bf 27}, 215010 (2010)
  [arXiv:1002.4826 [gr-qc]].

\bibitem{Godazgar:2010ks}
  M.~Godazgar,
  Class.\ Quant.\ Grav.\  {\bf 27}, 245013 (2010).
  [arXiv:1008.2955 [gr-qc]].

\bibitem{DeSmet:2003kt}
  P.~-J.~De Smet,
  Gen.\ Rel.\ Grav.\  {\bf 36}, 1501-1504 (2004).
  [gr-qc/0312021].

\bibitem{DeSmet:2003bt}
  P.~-J.~De Smet,
  Class.\ Quant.\ Grav.\  {\bf 20}, 2541-2552 (2003).
  [gr-qc/0302081].

\bibitem{hacyan}
S. Hacyan, Phys. Lett. A75, 23 (1979).

\bibitem{Karlhede:1985ds}
  A.~Karlhede,
  Class.\ Quant.\ Grav.\  {\bf 3}, L1 (1986).

\bibitem{Gauntlett:2002nw}
  J.~P.~Gauntlett, J.~B.~Gutowski, C.~M.~Hull, S.~Pakis and H.~S.~Reall,
  Class.\ Quant.\ Grav.\  {\bf 20}, 4587 (2003)
  [arXiv:hep-th/0209114].

\bibitem{Gross:1983hb}
  D.~J.~Gross and M.~J.~Perry,
  Nucl.\ Phys.\  B {\bf 226}, 29 (1983).

\bibitem{Sorkin:1983ns}
  R.~d.~Sorkin,
  Phys.\ Rev.\ Lett.\  {\bf 51}, 87 (1983).

\bibitem{Coley:2009is}
  A.~Coley and S.~Hervik,
  Class.\ Quant.\ Grav.\  {\bf 27}, 015002 (2010)
  [arXiv:0909.1160 [gr-qc]].


\bibitem{Krtous:2008tb}
  P.~Krtous, V.~P.~Frolov and D.~Kubiznak,
  Phys.\ Rev.\  D {\bf 78}, 064022 (2008)
  [arXiv:0804.4705 [hep-th]].


\bibitem{Mason:2008ih}
  L.~Mason and A.~Taghavi-Chabert,
  J.\ Geom.\ Phys.\  {\bf 60}, 907 (2010)
  [arXiv:0805.3756 [math.DG]].

\bibitem{Houri:2007xz}
  T.~Houri, T.~Oota and Y.~Yasui,
  Phys.\ Lett.\  B {\bf 656}, 214 (2007)
  [arXiv:0708.1368 [hep-th]].




\bibitem{Hughston:1988nz}
  L.~P.~Hughston and L.~J.~Mason,
  Class.\ Quant.\ Grav.\  {\bf 5}, 275 (1988).

\bibitem{trautman}
P. Nurowski and A. Trautman, Diff. Geom. Appl. {\bf 17}, 175 (2002). 

\bibitem{TaghaviChabert:2010bm}
  A.~Taghavi-Chabert,
  arXiv:1011.6168 [gr-qc].
\end{thebibliography}
\end{document}